\documentstyle[12pt]{article}
\begin{document}
\setcounter{page}{0}
\thispagestyle{empty}
\def\Journal#1#2#3#4{{#1} {\bf #2}, #3 (#4)}
\def\NCA{\em Nuovo Cimento}
\def\NIM{\em Nucl. Instrum. Methods}
\def\NIMA{{\em Nucl. Instrum. Methods} A}
\def\NPA{{\em Nucl. Phys.} A}
\def\NPB{{\em Nucl. Phys.} B}
\def\PLB{{\em Phys. Lett.}  B}
\def\PRL{\em Phys. Rev. Lett.}
\def\PRC{{\em Phys. Rev.} C}
\def\PRD{{\em Phys. Rev.} D}
\def\ZPC{{\em Z. Phys.} C}

\centerline
{\large \bf What Can We Learn from Transverse Momentum Spectra}
\medskip
\centerline
{\large \bf in Relativistic Heavy Ion Collisions?}
\vskip 1cm

\centerline{\bf Jan-e~Alam$^1$, J. Cleymans$^2$,  K.~Redlich$^{3,4}$,
H.~Satz$^5$ }
\vskip 1cm

$^1$Variable Energy Cyclotron Centre, 1/AF Bidhan Nagar, Calcutta
6700 064, India \par
$^2$Department of Physics, University of Cape Town, Rondebosch 7700,\
South Africa  \par
$^3$Gesellschaft f\"ur Schwerionenforschung (GSI),
D-64220 Darmstadt, Germany\par
$^4$On leave from: Department of Theoretical Physics, University of
Wroclaw, Poland\par
$^5$Fakult\"at f\"ur Physik, Universit\"at Bielefeld,
D-33501 Bielefeld, Germany\par
\vskip 1.5cm

\begin{abstract}

We compare the predictions of different models for transverse
momentum spectra in relativistic heavy ion collisions. Particular
emphasis is given to hydrodynamic flow with different assumptions for
the radial expansion, and to models based on a superposition of
fireballs along the transverse direction (random walk). It is shown
that present experimental data cannot distinguish between the
models under consideration.
\end{abstract}

\vskip 1cm

\bigskip

\noindent{\large \bf 1.\ Introduction}

\bigskip

Recent results on particle yields in $Pb-Pb$ collisions at CERN
\cite{NA44,NA49} show that the average transverse momenta of hadronic
secondaries are considerably larger than those in $p-p$ collisions, and
that the increase depends strongly on the mass of the hadron.
One interpretation of this effect is that the observed
hadrons are subject to transverse flow~\cite{heinz}; this is a velocity
effect and therefore heavy particles acquire more momentum than light
particles. Recently, however, renewed attention has been
drawn to the fact that there is a ``normal'' $p_t$-broadening observed
in all reactions involving nuclear targets, from Drell-Yan dilepton
production to low $p_t$ mesons or baryons \cite{leonidov,jaipur}.
For high $p_t$ hadrons this is generally referred to as Cronin
effect \cite{cronin}. Here, as well as in Drell-Yan dilepton or in
quarkonium production, it is accounted for by the fact that successive
parton scatterings rotate the collision axis relative to the beam axis:
any given transverse momentum distribution will appear broadened when
it is measured in the reference frame fixed by the incident primary
beams.

These considerations were applied to low $p_t$ hadron production in
nuclear collisions~\cite{leonidov}, assuming that successive collisions
in nuclear reactions lead to a random walk in the transverse rapidity
plane. The displacement per collision in transverse rapidity was
determined from $p-A$ collisions; the normalized $p_t$ spectra for $A-B$
collisions are then predicted parameter-free and agree quite well with
the mentioned data from $Pb-Pb$ interactions~\cite{NA44,NA49}. In
particular, this ``normal'' $p_t$-broadening also reproduces the
increase with increasing hadron mass, giving more broadening for
nucleons than for kaons, and more for kaons than for pions. A very
recent study~\cite{kapusta} has gone even further and determined the
``kick per collision'' from $p-p$ rather than from $p-A$ data.

The  $p_t$-broadening observed in nuclear reactions can thus be quite
well accounted for by random walk collision axis rotations. Perhaps one
might consider such a phenomenon as a precursor for
``transverse flow''. Nevertheless, any hydrodynamic description of
$p_t$-distributions from $A-B$ collisions, with the flow velocity as open
parameter, has to face two  questions: why is there also
broadening in $p-A$ interactions? and why can the ``flow velocity'' in a random
walk approach be determined from $p-A$ or even from $p-p$ collisions? Perhaps
only two-particle correlations, rather than single particle spectra, can
distinguish between hydrodynamic flow and a random walk
approach~\cite{esumi}.

In this paper we want to show explicitly that the present information on
transverse momentum spectra can be adequately explained in conceptually
different models and thus cannot determine the origin of the observed
broadening.

We will first consider a hydrodynamic model with transverse flow and
rapidity scaling. In this case, all spectra show a characteristic
flattening for very small values of $p_t$. This is due to the flow
which adds momentum to very slow particles and hence depletes this
kinematic region. Next, we will study two simplified forms for
the transverse velocity: a constant velocity and one with a polynomial
dependence on the transverse radius. All three versions are able to
reproduce the data, with some caveats to be discussed later on in
the paper. In particular, the degree of flattening characteristic of a
full hydrodynamic picture does not seem to be present in the data. As
alternative description, we consider the behaviour obtained from the
random walk model of~\cite{leonidov}. In this case, one typically
overshoots the data somewhat at low transverse momenta. This could
well be due to the gaussian distribution used to describe the
successive kicks of the projectile; it has the tendency to accumulate
particles around the origin, and an exponential distribution might lead
to a better description of the present data. Once the quality of the
data improves, such a fit should definitely be made; in the present
paper, we restrict ourselves to the models so far proposed in the
literature. An inherent difficulty of the random walk model is
that each one of the successive collisions is determined by results
obtained from $p-A$ collisions, so that in particular the relative
particle abundances will also be the same as in $p-A$ collisions.
There is therefore no mechanism to obtain the observed increase in
the abundance of strange particles.

\bigskip

\noindent {\large \bf 2.\ Hydrodynamic Expansion}

\bigskip

The   momentum   distribution  of  particles  is  given  by  the
well-known Cooper-Frye formula~\cite{cooper}
\begin{equation}
E{dN\over  d^3p}  =  {g\over  (2\pi)^3}\int_{\sigma}  f(x,p)p^\mu
d\sigma_\mu,
\end{equation}
where the integration has to be performed over the freeze-out surface
described by $\sigma_\mu$.
For the temperatures under consideration it is safe to neglect quantum
statistics and we will therefore work with the Boltzmann distribution
from now on; the generalization to Fermi-Dirac or Bose-Einstein
statistics is straight-forward. We thus have
\begin{equation}
f(x,p) = \exp [(-p.u+\mu)/T],
\end{equation}
where $T$, $\mu$ and $u^\mu$ are the (space-time dependent) temperature,
  chemical   potential  and
four-velocity of the fireball, respectively. For a fireball at rest we
have
\begin{equation}
u^\mu = (1,\vec{0}),
\end{equation}
for one boosted in, e.g., the $x$-direction,
\begin{equation}
u^\mu = (\cosh y_T, \sinh y_T ,0,0),
\end{equation}
while for a boost in the azimuthal direction $\phi$ we get
\begin{equation}
u^\mu = (\cosh y_T, \cos\phi\sinh y_T ,\sin\phi\sinh y_T,0).
\end{equation}
For a static fireball undergoing instantaneous freeze-out, the
corresponding freeze-out surface is given by
\begin{equation}
d\sigma^\mu = (d^3x,\vec{0}),
\end{equation}
so that we obtain the standard expression (in Boltzmann approximation)
\begin{equation}
E{dN\over  d^3p}  =  {gV\over  (2\pi)^3}E \exp (-E/T).
\end{equation}
For boost-invariant cylindrical expansion along the
$z$-axis (recall $d^4x = \tau d\tau dy rdrd\phi $), we have
\begin{equation}
d\sigma^\mu = (\tau dy rdrd\phi,\tau d\tau dy rd\phi),
\end{equation}
where the second component is in the $\hat{r}$ direction, i.e.
perpendicular to the surface of the cylinder. For the case where the
flow is azimuthally symmetric, i.e.  when an  average  is  made  over
all  events  or when only head-on collisions are considered, one has
therefore~\cite{ruuskanen,blaizot}
\begin{eqnarray}
\left( {dN\over dy m_tdm_t}\right)_{y=0}
&=& {g\over \pi} \int_\sigma r~dr~\tau_F(r)    \nonumber\\
& & \left\{ m_tI_0 \left( {p_t\sinh y_t\over T} \right)\right.
          K_1 \left( {m_t\cosh y_t\over T} \right) \nonumber \\
& & -\left( {\partial\tau_F\over\partial r} \right) p_t
          I_1 \left( {p_t\sinh y_t\over T} \right)
    \left. K_0 \left( {p_t\cosh y_t\over T} \right) \right\}
\end{eqnarray}
where $\tau_F(r)$ refers to the freeze-out time which in general depends
on $r$, so that the center of the cylinder freezes out before the
surface.

We have taken the values for the derivative $\partial\tau_F(r)/\partial
r$ from reference~\cite{crs1}; in the figures, the results based on
eq. (9) are labeled {\it ``hydro 2+1 dim''}.

\bigskip

\noindent{\large \bf 3.\ Transverse Flow}

\bigskip

If the space-time development typical of the hydrodynamic expansion is
not taken into account and one instead considers bubbles of fluid
receiving boosts in the transverse direction, then it is natural to
take the freeze-out time as independent of $r$,
\begin{equation}
\tau_F(r) = \tau_F
\end{equation}
so that  everywhere within the volume the particles freeze out
simultaneously and the second term in equation (9) disappears. One is
then left with
\begin{equation}
\left( {dN\over dy m_tdm_t}\right)_{y=0}
= {gV \over \pi^2} \int_0^1 \xi~d\xi~
  m_tI_0 \left( {p_t\sinh y_t\over T} \right)
          K_1 \left( {m_t\cosh y_t\over T} \right)
\end{equation}
where $V$ is defined to be $\pi R_F^2\tau_F$, where $R_F$ denotes
the value of the radius at freeze-out. The variable $\xi = r/R_F$.
is directly related to $r$, measuring the distance from the axis of
the cylinder. A  natural possibility is to allow for a velocity
profile of the form~\cite{heinz,kaempfer,PBM}
\begin{equation}
\tanh y_t = v_t=v_{\perp}^{aver} {\alpha +2\over 2}\left( {r\over
R_A}\right)^\alpha.
\end{equation}
In the figures we will label the results based on  equation (11)
(with $\alpha=1$) as ``$v_t$-profile''.

If the transverse expansion is taken to be independent of $r$,
so that the expansion velocity is the same in the center as it is
on the surface, then one obtains
\begin{equation}
\left( {dN\over dy m_tdm_t}\right)_{y=0}
= {gV\over 2\pi^2}   m_tI_0 \left( {p_t\sinh y_t\over T} \right)
          K_1 \left( {m_t\cosh y_t\over T} \right),
\end{equation}
where the volume $V$ is again determined by $V=\pi R_F^2\tau_F$. It should be
noted that a constant transverse velocity leads to a problem at the
origin, since the transverse velocity must be zero at this point
because of symmetry considerations. In the figures we will refer to the
results of eq. (13) as {\it ``$v_t$=const''}.

\bigskip

\noindent{\large \bf 4.\ Random Walk}

\bigskip

In the random walk approach, nuclear collisions are assumed to be much
like elementary $p-p$ collisions, except that in the successive
scatterings occurring in nuclear targets, the collision axis will be
rotated.

It was shown recently \cite{becattini} that a thermal description
accounts quite well for the particle ratios in $e^+-e^-$, $p-p$ and
$\bar{p}-p$ collisions. We therefore follow this picture and assume
that in each of the successive interactions of a nuclear collision one
creates a fireball just like that formed in a nucleon-nucleon
collision; there is no need to introduce hydrodynamic flow for such
little fireballs. The only difference now is that after the first
collision the next one will generally occur at some non-vanishing
transverse velocity. We therefore have to know the propagation of
transverse momentum through successive collisions. It seems simplest to
assume that this will follow a random walk pattern, which in
\cite{leonidov} was taken to be Gaussian. As mentioned, one should
eventually study different distributions in such an approach; we will
restrict ourselves to the Gaussian
\begin{equation}
f_{pA}(\rho ) = \left[ {4\over \pi\delta^2_{pA}}\right]^{1/2}\exp
(-\rho^2/\delta^2_{pA}),
\end{equation}
where $\rho$ is the transverse rapidity and
\begin{equation}
\delta_{pA}^2 = (N_A-1)\delta^2,
\end{equation}
denotes the kick per collision $\delta$ as determined from $p-A$
interactions. The corresponding distribution for an $A-B$ collision can
then be obtained by the convolution
\begin{equation}
f_{AB}(\rho)=\int f_{pA}(\rho ') f_{pB}(\rho '')\delta(\rho ' +
\rho '' -\rho)d\rho ' d\rho '',
\end{equation}
leading to
\begin{equation}
f_{AB}(\rho ) = \left[ {4\over \pi\delta^2_{AB}}\right]^{1/2}\exp
(-\rho^2/\delta^2_{AB}),
\end{equation}
with
\begin{equation}
\delta_{AB}^2 = (N_A+N_B-2)\delta^2.
\end{equation}
For simplicity  we  have followed the analysis of \cite{leonidov} and
have taken all fireballs at the same temperature; again this can easily
be generalized to a more general situation with a distribution in
temperature. The final expression thus becomes here
\begin{equation}
\left( {dN\over dy m_tdm_t}\right)_{y=0}
= {gV\over 2\pi^2}
  \left[ {4\over \pi\delta^2_{AB}}\right]^{1/2}\int d\rho\exp
(-\rho^2/\delta^2_{AB})
m_tI_0 \left( {p_t\sinh\rho\over T} \right)
          K_1 \left( {m_t\cosh\rho\over T} \right).
\end{equation}
It should be noted that the volume in the eq.\ (19) refers to the
volume of the system as observed in a $p-p$ collision, since each
collision in the random walk produces a $p-p$ type of fireball. For
normalized distributions, it will of course drop out.

If we now introduce a boost-invariant distribution of fireballs along
the longitudinal rapi\-di\-ty axis, we finally obtain by integrating
over the fireball distributions
\begin{eqnarray}
\left( {dN\over dy m_tdm_t}\right)_{y=0}
&=& {gV\over 2\pi^2}
  \left[ {4\over \pi\delta^2_{AB}}\right]^{1/2}\int d\rho\exp
(-\rho^2/\delta^2_{AB})\nonumber\\
& & \int_{-Y_L}^{Y_L} dY
m_t\cosh YI_0 \left( {p_t\sinh\rho\over T} \right)
          K_1 \left( {m_t\cosh Y\cosh\rho\over T} \right).
\end{eqnarray}
In the figures the results based on eq. (20) will be denoted by
{\it``random walk''}.

\bigskip

\noindent{\large \bf 5.\ Results}

\bigskip

We now want to compare the models described in the previous sections
with the latest experimental data from  $Pb-Pb$ collisions obtained by
the NA44~\cite{NA44} and the NA49~\cite{NA49} collaborations at CERN.
These data were obtained in different kinematic regions. In the
figures, we have simply scaled the NA44 data to the NA49 data. The
parameters used in the different descriptions are listed in Table 1.

In Fig.\ 1, we show the NA49 data for negative hadrons together with
the NA44 results for $\pi^+$. All models show good agreement with the
NA49 data, but they cannot reproduce the steeper NA44 $\pi^+$ behaviour
at small $p_t$ together with that in the larger NA49 $p_t$ range.

In Fig.\ 2, we show the NA49 data for the surplus of positive
hadrons, $h^+-h^-$, and the NA44 data for protons. In this case there
is good agreement between both data sets and all models, except that
random walk overshoots and the full hydrodynamic model undershoots the
behaviour at very low $p_t$. As noted, in the random walk picture,
the choice of another distribution in the ``kick''-parameter $\delta$
may well bring the spectrum down. The characteristic low $p_t$ dip in
the hydrodynamical model is reduced somewhat when we choose a different
parameterization of the transverse velocity.

In Fig.\ 3, we show the NA49 data for neutral hyperons, i.e. $\Lambda$
and $\Sigma^0$. In this case, the hydrodynamical description with
constant $v_t$ does quite poorly, while all other models give a
reasonable description of the data.

In Fig.\ 4, we show the NA49 data for $K_s^0$ and the NA44 results for
$K^+$. There is again good agreement between data and models, except
in the low $p_t$ region, where the full hydrodynamic model shows a
depletion not present in the measured yield.

%
%
%
\begin{table}[h]
\begin{center}
\begin{tabular}{|c|c|c|c|} \hline
\hline
  & & & \\
Model                &  T$_f$ [GeV] & $<v_\perp >$ & References    \\
                     &           &       &          \\
\hline
$v_t=const$          & 0.15     &   0.37   &     \\
$v_t-profile$    & 0.12    & 0.43   & [12]        \\
{\sl random walk}          &  0.15     &   -   &  [6]    \\
{\sl hydro 2+1 dim}           &  0.115    &  0.5&     \\
hydrodynamics (\cite{sollfrank-hydro}) &  0.14     &   0.34   
&  \cite{sollfrank-hydro}\\
hydrodynamics (\cite{ornik})  &  0.143    &         &  \cite{ornik} \\
from particle ratio  &  0.17 - 0.19   &   -      & [14,15,17,18]  \\
\hline
\end{tabular}
\end{center}
\caption{Results for freeze-out parameters in different models.}
\end{table}

Finally we show in Fig.\ 5 how the different fits tend to disperse
for larger values of the transverse momenta. A similar behavior is
obtained for all particle spectra.

\bigskip

\noindent{\large \bf 6.\ Conclusions}

\bigskip

We have compared the latest experimental results on transverse momentum
spectra in $Pb-Pb$ collisions with several different models. All of
them manage to describe these data reasonably well, apart from some low
$p_t$ discrepancies; hence transverse spectra so far really do not allow
us to decide between a hydrodynamic or random walk origin of the
observed broadening.

Moreover, several further problems remain unanswered: all models
call for a rather low freeze-out temperature, as seen in Table 1, where
we have for comparison also listed results 
obtained in refs.~\cite{sollfrank-hydro,ornik}. Our results are typically in
the range $T_F\sim 120-150$ MeV. This is low when compared to the
temperatures needed to explain the hadron abundances in particle ratios,
where one needs values in the range of 170-190 MeV but it must be noted 
of course that the
various hydrodynamic models do not take into account 
resonance decays. The random walk
model
describes the data quite well but it will lead to the same hadronic
abundances as those observed in $p-A$ collisions. Thus it cannot
reproduce the increase in strangeness production reported by all
experimental groups. Besides the uncertainty about the origin of the
observed transverse broadening, we thus also still lack a
consistent account of both slopes and particle abundances.

\bigskip

\noindent{\large \bf Acknowledgments}

\bigskip

We thank many colleagues in the heavy ion community for stimulating
discussions, in particular U.\ Heinz, J.\ Sollfrank and D.\ K.\
Srivastava. K.R. acknowledges the support of the GSI Darmstadt and of
the KBN 2-P03B-09908, J.C. thanks the URC of the University of Cape
Town and the FRD (Pretoria) for financial support and the GSI Darmstadt
for   hospitality.   J.A.   is   grateful   to   the   Deutsche
Forschungsgemeinschaft  (DFG)  and  the Indian National Science
Academy (INSA) for financial support of this work through their
bilateral exchange program.

%
%
%
{\bf Figure Captions}\\
\begin{description}
\item{Figure 1 :} Comparison between model predictions and the
data for pion distributions.
\item{Figure 2 :} Comparison between model predictions and data
for proton distributions.
\item{Figure 3 :} Comparison between model predictions and data
for the $\Lambda$ distribution. 
\item{Figure 4 :} Comparison between model predictions and
data for kaon distributions. 
\item{Figure 5 :} Comparison between model predictions and
data for the deuteron distribution.
\end{description}
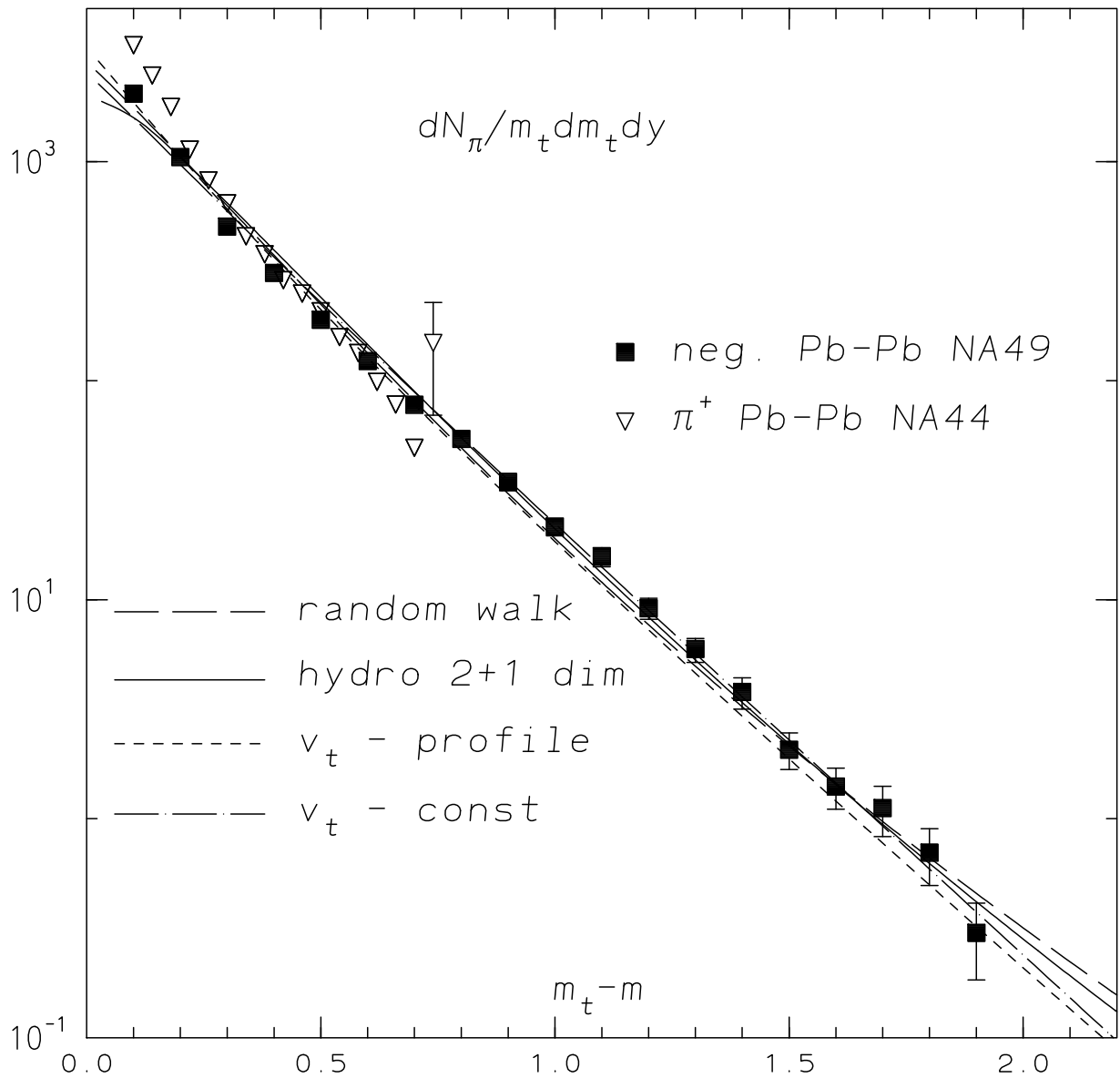
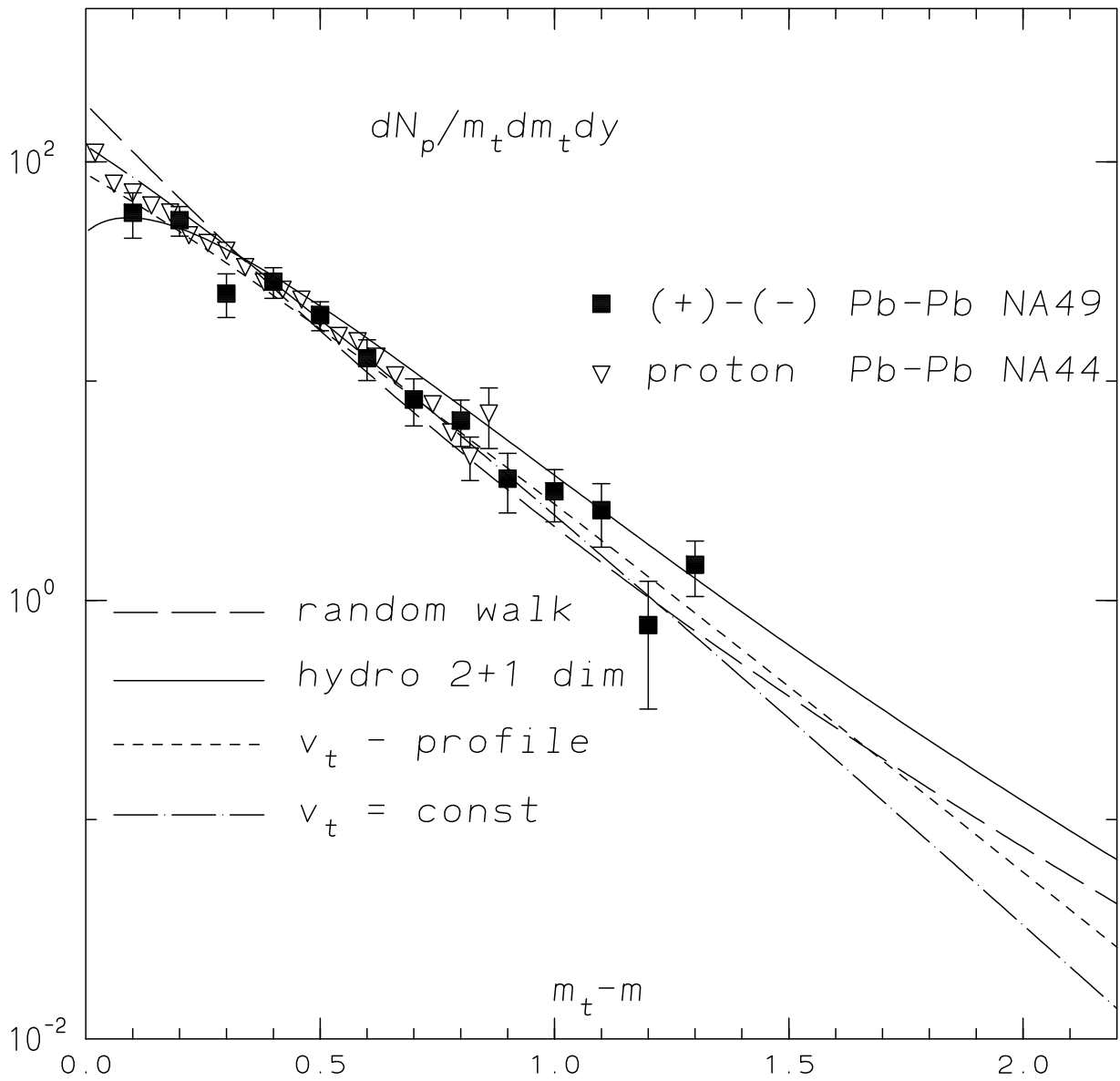
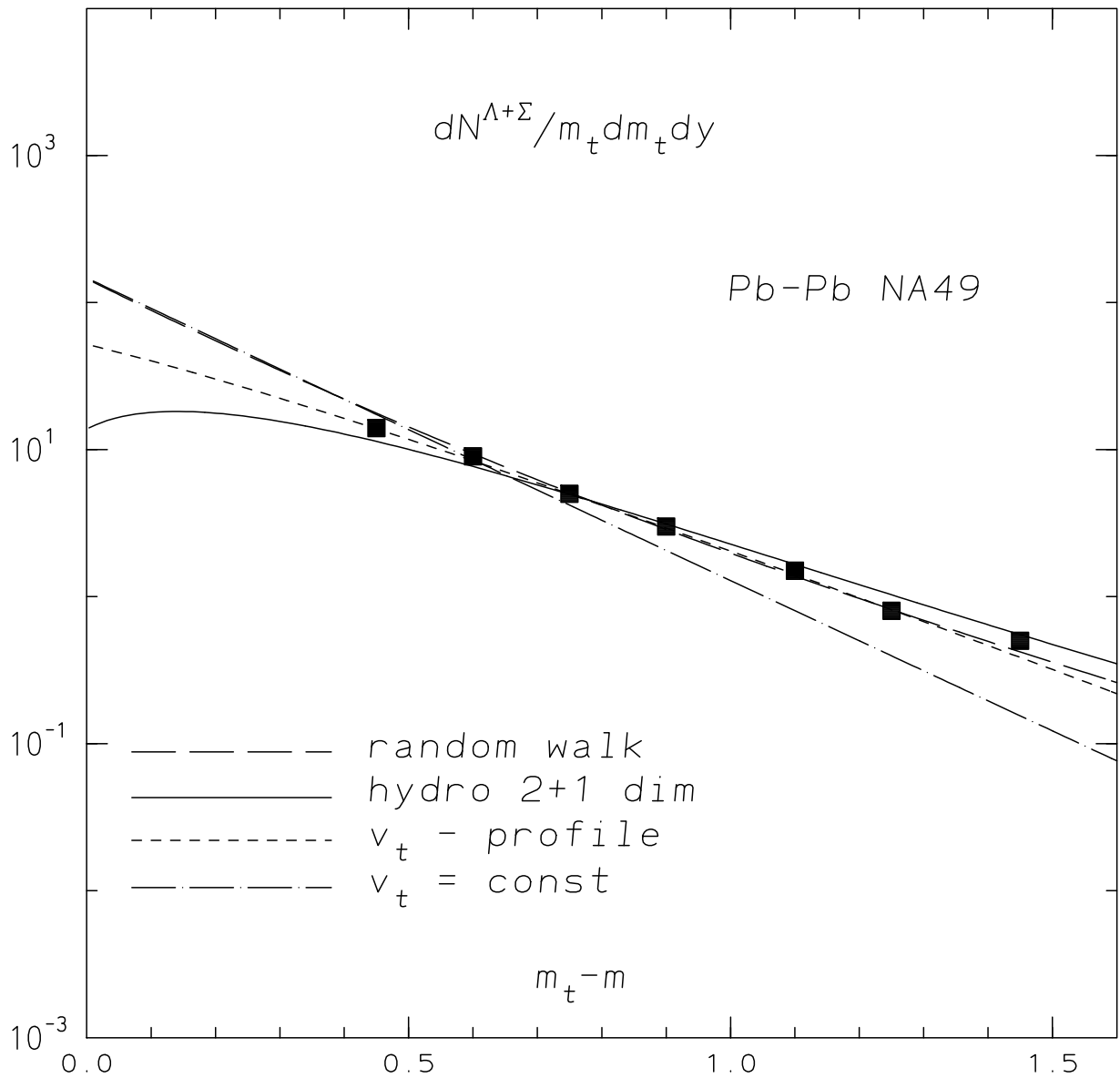
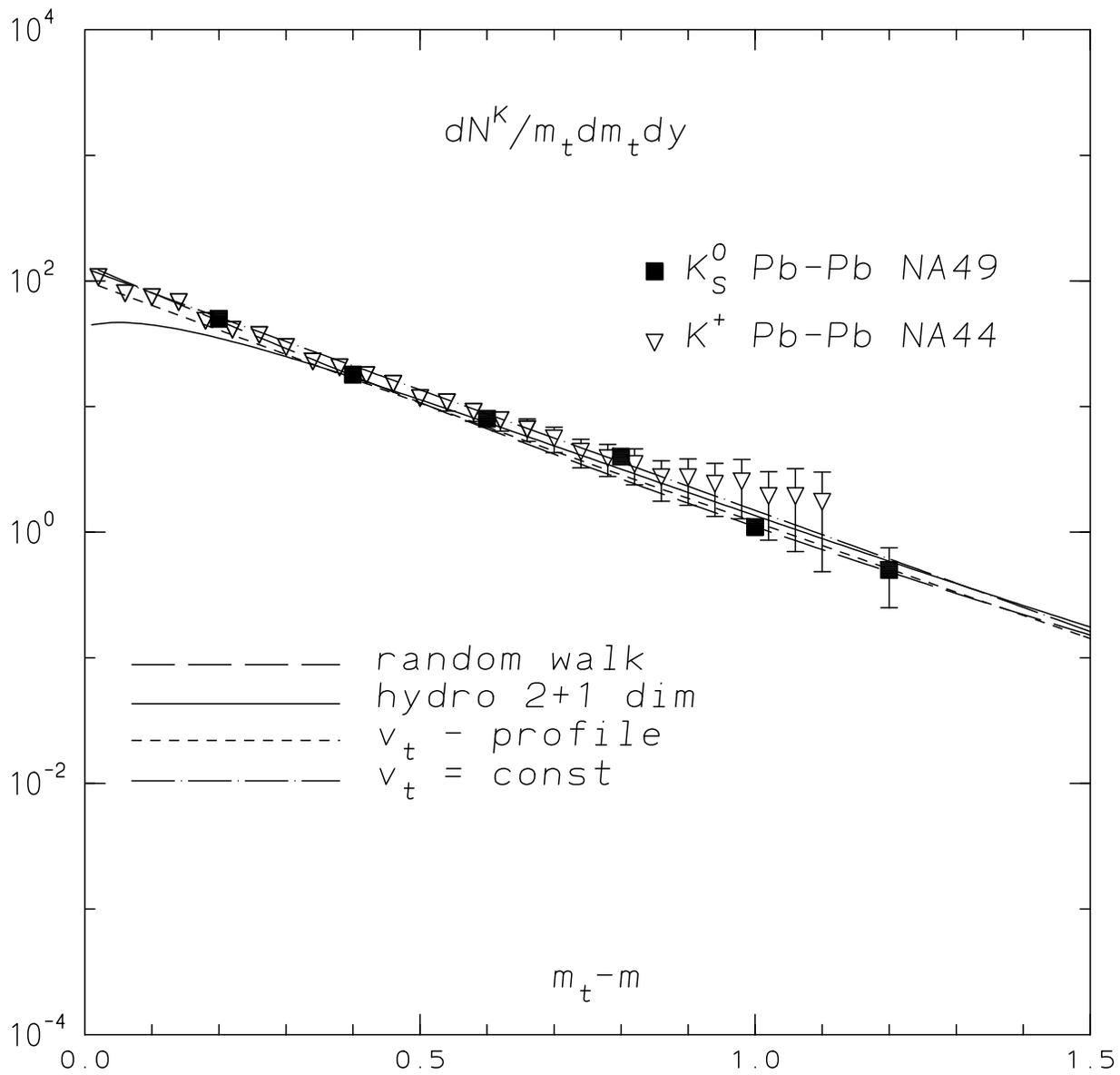
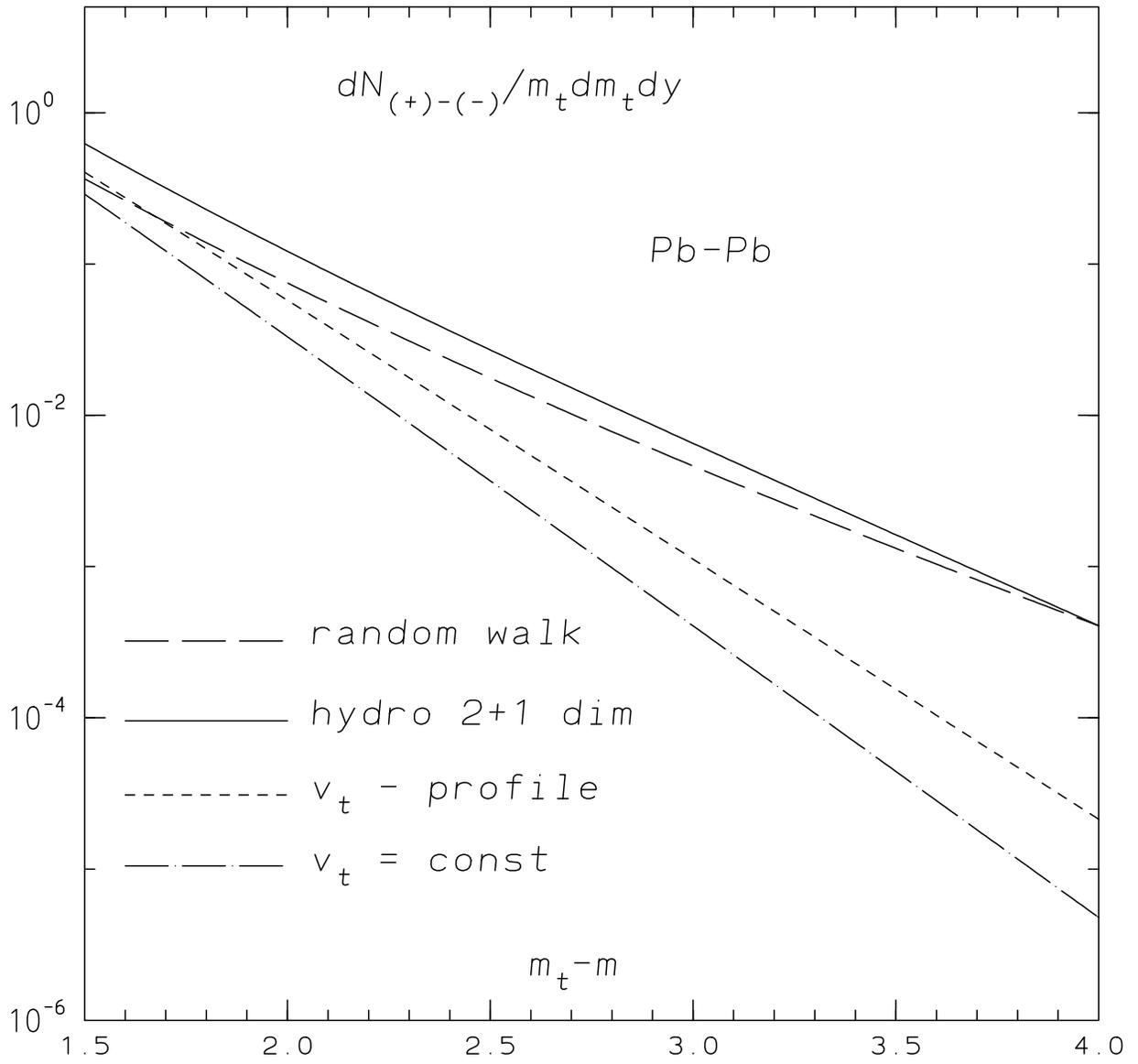
\end{document}